\documentclass[conference]{IEEEtran}
\IEEEoverridecommandlockouts
\usepackage{cite}
\usepackage{amsmath, amssymb, amsfonts}
\usepackage{tipa}
\usepackage{algorithm}
\usepackage{algpseudocode}
\usepackage{graphicx}
\usepackage{xcolor}
\usepackage{dblfloatfix}
\usepackage{textcomp}
\usepackage{xcolor}
\usepackage{authblk}
\usepackage{hyperref}
\usepackage{orcidlink}

\def\BibTeX{{\rm B\kern-.05em{\sc i\kern-.025em b}\kern-.08em
    T\kern-.1667em\lower.7ex\hbox{E}\kern-.125emX}}
\begin{document}

\title{The OCON model: an old but green solution for distributable supervised classification for acoustic monitoring in smart cities\\
\thanks{This work was partially supported by the EU under the Italian NRRP of NextGenerationEU, partnership on “Telecommunications of the Future” (PE00000001 - program “RESTART”)}
}

\author[1]{Stefano Giacomelli \orcidlink{0009-0009-0438-1748}}
\author[1]{Marco Giordano \orcidlink{0009-0001-1649-6085}}
\author[2]{Claudia Rinaldi \orcidlink{0000-0002-1356-8151}}

\affil[1]{DISIM - University of L'Aquila, L'Aquila, Italy}
\affil[2]{CNIT - Consorzio Nazionale Interuniversitario per le Telecomunicazioni, L'Aquila, Italy}

\maketitle
\thispagestyle{plain}
\pagestyle{plain}
\begin{abstract}
This paper explores a structured application of the \textit{One-Class} approach and the \textit{One-Class-One-Network} model for supervised classification tasks, focusing on \textit{vowel phonemes classification} and \textit{speakers recognition} for the Automatic Speech Recognition (ASR) domain. For our case-study, the ASR model runs on a proprietary sensing and lightning system, exploited to monitor acoustic and air pollution on urban streets. We formalize combinations of \textit{pseudo}-Neural Architecture Search and Hyper-Parameters Tuning experiments, using an \textit{informed} grid-search methodology, to achieve classification accuracy comparable to nowadays most complex architectures, delving into the speaker recognition and energy efficiency aspects. Despite its simplicity, our model proposal has a very good chance to generalize the language and speaker genders context for widespread applicability in computational constrained contexts, proved by relevant statistical and performance metrics. Our experiments code is openly accessible on our \href{https://github.com/StefanoGiacomelli/Vowel\_phonemes\_Analysis\_and\_Classification\_by\_means\_of\_OCON\_rectifiers\_Deep\_Learning\_Architectures/tree/main}{GitHub}.
\end{abstract}


\section{Introduction}
Acoustic sensing for the safety, security, and monitoring of urban and non-urban environments is becoming increasingly important. This trend is driven by the widespread adoption of the smart cities vision by Western municipalities \cite{alias2019, pastor2020} and the need to protect ecosystems in wild areas \cite{farina_review, Markolf2022TowardPA, johnson2023}. Solutions proposed in the literature over the years addressed two critical topics: communication networks and Machine Learning (ML) \cite{jaume2021, Gouvea_2023}. Collected data need to be transmitted and processed, with the sequence of these operations depending on many factors (not addressed in this work). Recent technological advancements in both fields are worth exploring and new communication networks like $5$G and $6$G enable previously impossible applications by introducing new concepts such as network slicing, network functions virtualization, orchestration, multi-access edge computing, Open Radio Access Networks (O-RAN), and software-defined networking \cite{rinaldi2021exploitation, marotra2023oran}. Additionally, ML and Neural Networks (NNs) are extensively spreading across various fields, being highly desirable for acoustic monitoring due to their proved accuracy levels \cite{RENAUD2023119568, MARCINIUK20221087, informatics7030023, NIETOMORA2023e20275}.

\begin{figure}[!t]
\centering
\includegraphics[width=2.2in]{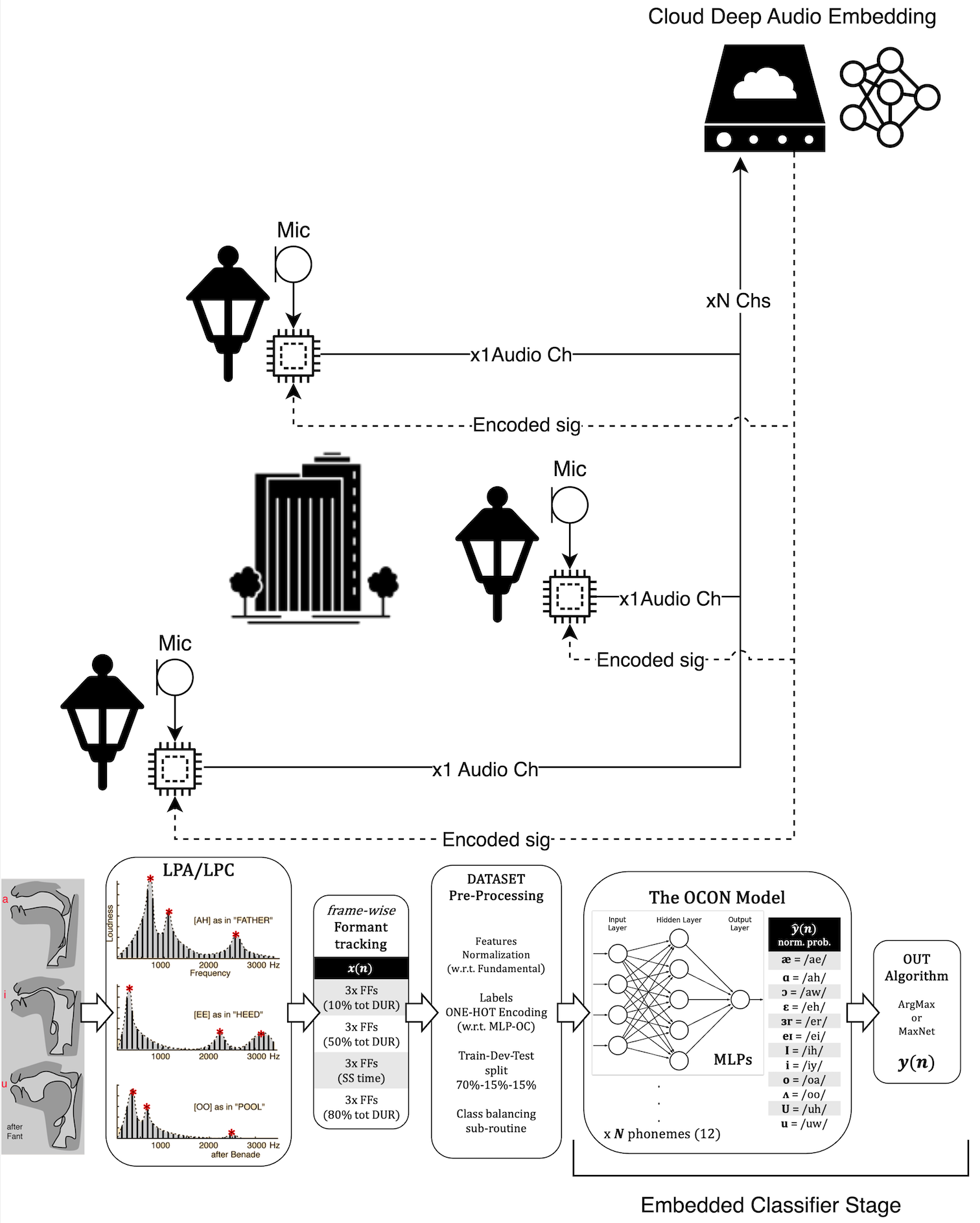}
\caption{Our smart city case-study scenario}
\label{smart_city}
\end{figure}

In a smart city scenario (Fig.\ref{smart_city}), key areas of interest are Sound Event Detection (SED) and Audio Tagging (AT), particularly for identifying the cause of exceeding a safe acoustic threshold. This operation can be performed either at the sensor location or remotely, especially when detailed information about the type of sound needs to be transmitted. This scenario fits perfectly within the field of the Internet of Sounds (IoS) \cite{turchet2023IoS}, which involves the interconnected network of devices capable of capturing, processing, and transmitting audio data with constrained computational resources \cite{MONDAL2022108252}. When adopting ML solutions for this purpose, care must be taken due to the limited computational capabilities of local sensing hardwares \cite{10335204, turchet_real_time}. One possible solution is to send all data to a remote cloud, another is to reduce the computational complexity of the NN to be used locally and improve its features abstraction capability, thus avoiding privacy issues too.

This paper evaluates a streamlined combination of Neural Architecture Search (NAS) and Hyper-Parameters Tuning (HPs-T) for designing abstraction/classification NNs models. We propose a modular ``One-Class One-Network'' (OCON) model, which consists of parallel binary classifiers (instead of a single multi-class layer) dedicated to simpler and specific SED tasks: phonetic and speakers recognitions \cite{vad_realtime}. By assessing data constraints and task complexities with respect to the current State-of-the-Art (SoA), we strive to develop a \textit{shallow} and optimizable sub-architecture, characterized by a \textit{sustainable} and straightforward re-training cycle (given the constrained computational and emission requirements). Moreover, we determine the minimum number of formant features required to achieve SoA accuracy levels in phonetic recognition.

Building on recent results \cite{Acce2406:OCON}, we also explore the opportunity of recognizing the gender of the speaker for enhanced contextual understanding, personalized response, and behavioral analysis. Recent progress in gender recognition has reached a SoA level employing advanced ML solutions, in conjunction with sophisticated signal pre-processing techniques, to achieve remarkable accuracy in identifying and interpreting vocal subtleties. These approaches encompass feature extraction in both temporal and (pseudo)-frequency domains \cite{hasan2021}, as well as NNs and Deep Learning (DL) model ensembles, which are now at the core of almost every Automatic Speech Recognition (ASR) system \cite{10.1007/978-981-19-0604-6_17, 10.1007/s11042-023-16438-y, gender_1, https://doi.org/10.1155/2022/4444388, MAVADDATI2024127429}.

\section{Methodologies}
We started by collecting a reliable audio \textit{dataset}, including multiple phonetic hues and gender diversity among speakers. We limited our linguistic research to the \textit{General American English} case-study, as defined by the International Phonetic Association (IPA). We decided to focus only on well-established pre-processed datasets (Sec. \ref{datasets}), designed by means of pre-arranged phraseological segments or specific words, like the \textit{/hVd/ containers} (where vowels are placed between an ``\textit{h}'' and a ``\textit{d}''). These segments were recorded, analyzed (formant analysis), and pre-processed to extract meaningful features (\textit{formant} frequencies) suitable for our NNs model, obtained following this steps:

(\emph{1}) segment speech signals into \textit{semantic frames}, either manually or automatically, following a pre-defined semantic grid (words/phonemes, and silences);

(\emph{2}) use Linear Predictive Coding/Analysis (LPC/LPA) to analyze isolated segments, obtaining a smoothed time-frequency aggregated spectral estimate (per audio frame);

(\emph{3}) extract the top $N$ spectral peaks using any \textit{peak estimation} algorithm, ensuring to track frame-by-frame continuities (contouring).

Additional post-processing is added to refine retrieved formant frequency tracks and create a suitable (\textit{features}) vector for the input layer of our NNs model. These pre-processing proves to be crucial in allowing the networks to learn related abstract representations effectively, thereby optimizing recognition accuracy.

\subsection{Datasets Review} 
\label{datasets}
As already discussed in \cite{Acce2406:OCON}, the choice of the dataset has been done basing on various reasoning such as phonetic complexity and gender balance, by analyzing in detail three of those freely available as the \textit{Peterson and Barney (PB)} \cite{10.1121/1.1906875}, the \textit{HGCW} database \cite{10.1121/1.411872} and the Texas Instruments \& Massachusetts Institute of Technology (TI-MIT) \textit{Corpus of Read Speech} \cite{10.35111/17gk-bn40}.

Recognizing the limitations of existing datasets for developing \textit{fluid} and robust generalizable solutions, we opted for the HGCW dataset, which offers the highest level of phonetic complexity (although being merely \textit{binary-labeled}, a known limitation for the speaker gender task). By leveraging pre-extracted formant data from the HGCW repository, we aim to expedite the data retrieval process, promote consistency with the literature, and streamline results evaluation.

\subsection{Features Pre-Processing \& Classification}
\label{features_sec}

The filename structure of the HGCW dataset (Table \ref{hgcw_filenames}) encodes essential phonetic and speaker features, which are crucial for a preliminary statistical analysis.
\begin{table}
\begin{center}
\caption{HGCW dataset filenames structure}
\label{hgcw_filenames}
\begin{tabular}{| c | c | c | c |}
\hline
\textbf{$1^{st}$ \textbf{character}}& {$2^{nd} \& 3^{rd}$ \textbf{ch.s}} & {$4^{th} \& 5^{th}$ \textbf{ch.s}} & \textbf{Example}\\
\hline
m (\textit{man}) & spk. n° ($50$ tot.) & ARPABet ch.s & \texttt{m10ae}\\
b (\textit{boy}) & / ($29$ tot.) & / & \texttt{b11ei}\\
w (\textit{woman}) & / ($50$ tot.) & / & \texttt{w49ih}\\
g (\textit{girl}) & / ($21$ tot.) & / & \texttt{g20oo}\\
\hline 
\end{tabular}
\end{center}
\end{table}

Pre-processing solutions applied have already been discussed in \cite{Acce2406:OCON}. We remark that the presence of \textit{null} features (in some samples) caused by authors algorithm failures, required further samples filtering, leading to additional under-representation of certain phoneme and speaker classes (Table \ref{hgcw_classes}), to maintain balance and thus learning consistency.
\begin{table}
\begin{center}
\caption{HGCW actual classes statistics}
\label{hgcw_classes}
\begin{tabular}{| c | c | c | c | c | c | c |}
\hline
\textbf{Phoneme} & \textbf{Samples} & \textbf{Boys} & \textbf{Girls} & \textbf{Men} & \textbf{Women} & \textbf{Label ID}\\
\hline
ae (\textbf{\textturnscripta}) & $134$ & $25$ & $17$ & $45$ & $47$ & $0$\\
ah (\textbf{a}) & $135$ & $24$ & $19$ & $45$ & $47$ & $1$\\
aw (\textbf{\textopeno}) & $133$ & $24$ & $18$ & $45$ & $46$ & $2$\\
eh (\textbf{\textepsilon}) & $139$ & $27$ & $19$ & $45$ & $48$ & $3$\\
er (\textbf{\textramshorns}) & $118$ & $26$ & $18$ & $37$ & $37$ & $4$\\
ei (\textbf{e}) & $126$ & $25$ & $17$ & $43$ & $41$ & $5$\\
ih (\textbf{y}) & $139$ & $27$ & $19$ & $45$ & $48$ & $6$\\
iy (\textbf{i}) & $124$ & $20$ & $18$ & $43$ & $43$ & $7$\\
oa (\textbf{o}) & $136$ & $25$ & $19$ & $45$ & $47$ & $8$\\
oo (\o) & $139$ & $27$ & $19$ & $45$ & $48$ & $9$\\
uh (\textbf{u}) & $138$ & $26$ & $19$ & $45$ & $48$ & $10$\\
uw (\textbf{\textturnm}) & $136$ & $25$ & $19$ & $44$ & $48$ & $11$\\
\hline
\textbf{TOTAL} & $1597$ & $301$ & $221$ & $527$ & $548$ & $12$\\
\hline 
\end{tabular}
\end{center}
\end{table}
Fundamental frequency tracks ($F0$) were retrieved by means of a $2$-way auto-correlation/zero-crossing pitch tracker, followed by a halving/doubling result evaluation sub-routine \cite{10.1044/jshr.3604.694}, while formant frequencies were estimated using LPA and peak retrieval with parabolic interpolation \cite{10.2307/3680788}. The resulting frequency trajectories were additionally refined with an interactive audio spectral editor, which was used for manual examination and interpolation of discontinuities.

We recall that different experimental sub-structures of the original dataset have been obtained, categorizing each sample relying on:

(\emph{1}) \textit{Phonemes grouping}, including $F0$ and the first $3$ formant frequencies at the \textit{steady state} (SS);

(\emph{2}) \textit{Speakers grouping} with the same features as above, relying on provided gender labels;

(\emph{3}) \textit{Phonemes grouping}, including $F0$ and a total of $12$ formant frequency values, with the first three formants sampled at $10\%$, $50\%$, SS, and $80\%$ of the total duration of the vowel nucleus.

To establish a consistent reference baseline, we analyzed classification algorithms evaluated on PB and/or HGCW dataset features (Table \ref{statistical_ML}) only. Linear Discriminant Analysis (LDA) \cite{10.1121/1.393381} and Generalized Linear Regression Models (GLM) \cite{10.21437/ICSLP.1992-167} resulted the most prominent and effective approaches. These methods were combined with innovative formant feature processing, such as the $3$\textit{D-auditory target zones} framework, using logarithmic formant distances \cite{10.1121/1.397862}. Other studies applied canonical auditory frequency transforms including the Bark scale \cite{10.1121/1.385079, 10.1121/1.385780}, Mel scale approximations \cite{10.1515/9783110873429}, and \textit{lin-to-log} frequency mapping \cite{Koening}.
\begin{table}
\begin{center}
\caption{Phonemes \& Speaker recognition w. PB dataset\\ (GM = \textit{geometric mean})}
\label{statistical_ML}
\begin{tabular}{| c | c | c | c | c |}
\hline
\textbf{Task} & \textbf{Data scale}& \textbf{Processing}& \textbf{ML}& \textbf{Accuracy}\\
\hline
Phoneme & Hz & \textit{Jacknife} & LDA & 81.8\%\\
& Log & None & GLM & 87.4\%\\
& / & -GM($F0$), $\cdot 0.333$& LDA & 86.3\%\\
& / & -($\bar{F1}, \bar{F2}, \bar{F3}$) & LDA & 89.5\%\\
& Bark & None & GLM & 86.2\%\\
& / & \textit{Jacknife} & LDA & 85.7\%\\
& / & -GM($F0$)& LDA & 85.3\%\\
& / & -($\bar{F1}, \bar{F2}, \bar{F3}$) & LDA & 88.3\%\\
& ERBs& None & GLM & 86.8\%\\
& / & -GM($F0$), $\cdot 0.5$& LDA & 87\%\\
& / & -($\bar{F1}, \bar{F2}, \bar{F3}$) & LDA & 88.8\%\\
\hline
Speaker & Hz & None & LDA & 89.6\%\\
& Bark & None & LDA & 88\%\\
& / & $\Delta F_n$ & LDA & 41.7\%\\
\hline
\end{tabular}
\end{center}
\end{table}

Research on phonetic NNs recognition has mainly focused on using LPA coefficients directly \cite{10.5120/7777-0862} or spectral/cepstral-derived features \cite{Kohonen}, often incorporating complex convolutional and/or recurrent modules. The only study on OCON phonetic classification \cite{10.21437/ICSLP.1998-40} reported improvements exclusively over TI-MIT data, using LPC features. Due to a limited comparative literature, we set a target average accuracy of $90$\%, aiming to improve results reported in \cite{10.1044/jshr.3604.694, 10.1121/1.411872} and \cite{Acce2406:OCON}.

\begin{figure}[!t]
\centering
\includegraphics[width=3.4in]{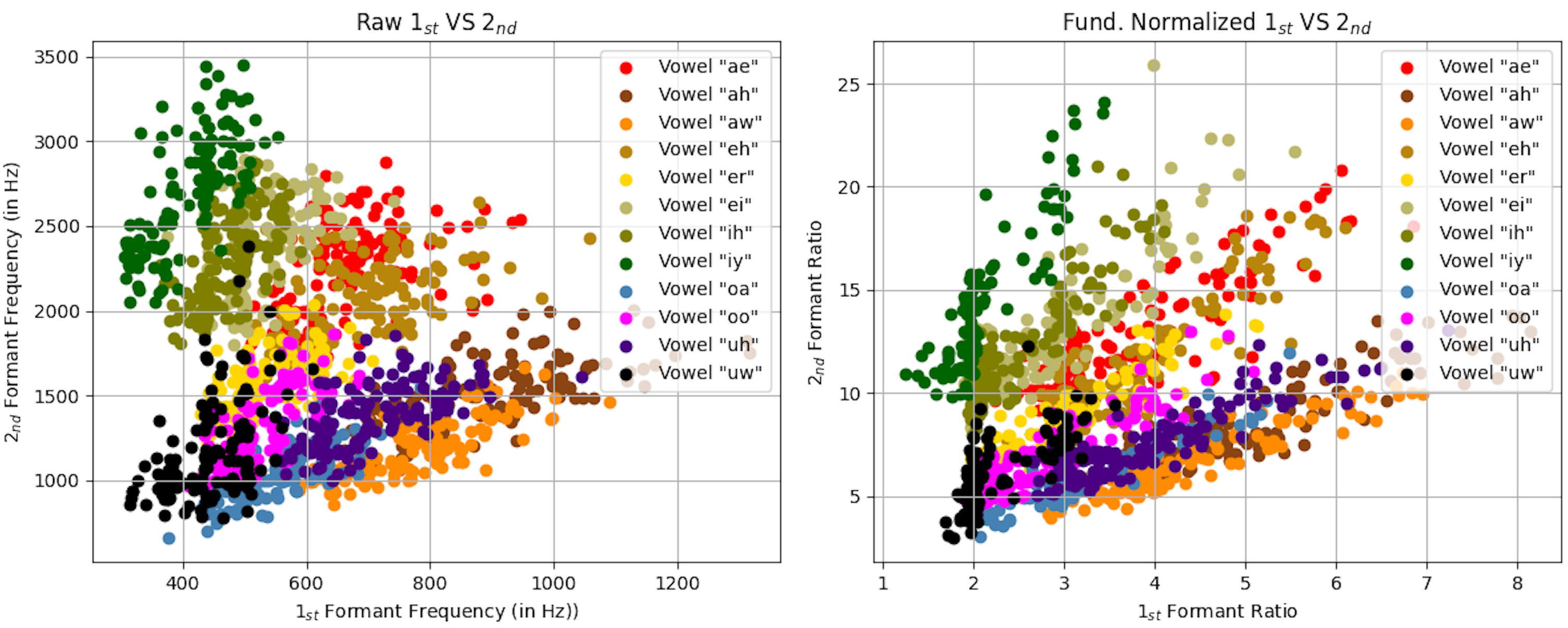}
\caption{HGCW dataset normalization ($2$D formantic projection)}
\label{hgcw_raw_vs_normalized}
\end{figure}

\begin{figure}[!t]
\centering
\includegraphics[width=3.5in]{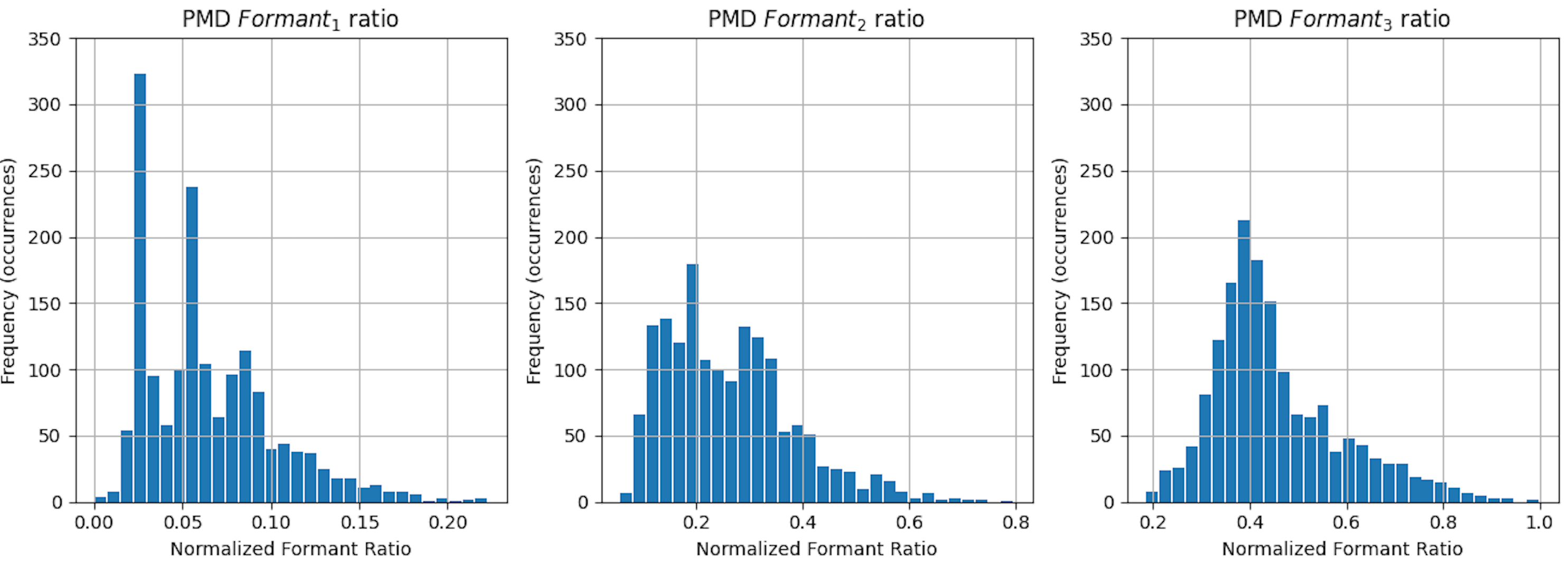}
\caption{HGCW Dataset PMDs (across all classes)}
\label{hgcw_pmds}
\end{figure}

Considering significant variations in $F0$ within speakers (due to physiological factors) and related pitch deviations caused by \textit{prosody}, we introduce a \textit{Linear Formant Normalization}, with respect to $F0$:
\begin{equation}
ratio(F_{i, n}) = \frac{F_{i, n}}{F0_n}
\end{equation}
where $F_i$ is the non-normalized $i^{th}$ formant (in Hz) and $F0_n$ is the fundamental frequency of the $n^{th}$ phoneme. Class distributing boundaries are improved, as shown in the $2$D formantic projection in Fig.\ref{hgcw_raw_vs_normalized}.
To enhance NNs training convergence, we applied \textit{min-max} scaling to normalize the entire feature set and we also examined Probability Mass Distributions (PMDs) of resulting formant ratios to assess the feasibility of \textit{Z-score} (standardization). However, the PMDs consistently exhibited a skewed distribution, resembling either Poissonian or Log-normal spread (Fig. \ref{hgcw_pmds}).

To preserve data resolution, we encoded all pre-processed features in a binary \texttt{NumPy} open-source compressed format (\texttt{.npz}), specifically designed to enhance data portability and re-usability.

\section{Practical Implementation}
In order to achieve both phonetic and speakers gender classification, we propose the exploitation of a specific OCON proposal, which models multi-output classification tasks using multiple independent exact-copies of the same optimized Multi-Layer Perceptron (MLP) reference architecture.

These configurations are derived through simplified and \textit{informed} NAS experiments (\textit{pseudo}-NAS) combined with HPs-T: in DL research, HPs-tuning involves optimizing architectural and learning parameters (such as layers, nodes, \textit{backpropagation} optimizers, learning rate etc.) to minimize the network cost function, between the predicted result (class) and the provided \textit{ground-truth} (label), in supervised learning contexts.

\subsection{Architecture \& Model}
\label{subsec:architecture_and_model}
MLPs, also referred to as \textit{Feed-forward} NNs or \textit{fully connected} (FC) layers, are essentially stacks of \textit{Perceptrons} (neurons) arranged in vertical layers (\textit{shallow} NNs), whose function is: 
\begin{equation}
y_n=\varphi\langle x, w_k\rangle = \varphi(x^\top w_k)=\varphi\left(\sum_{k=0}^{K}x_nw_k\right)
\end{equation}
where $x_n$ are the input features, $w_k$ a set of scaling coefficients (\textit{weights}) and $\varphi(\cdot)$ a non-linear ReLU function (\textit{activation}) \cite{10.1109/ICCV.2015.123}:
\begin{equation}
\varphi(x)=\max(0, x)=
    \begin{cases}
        x & \text{if } x > 0, \\
        0 & \text{otherwise}
    \end{cases}
\end{equation}

\begin{figure}[!t]
\centering
\includegraphics[width=3.3in]{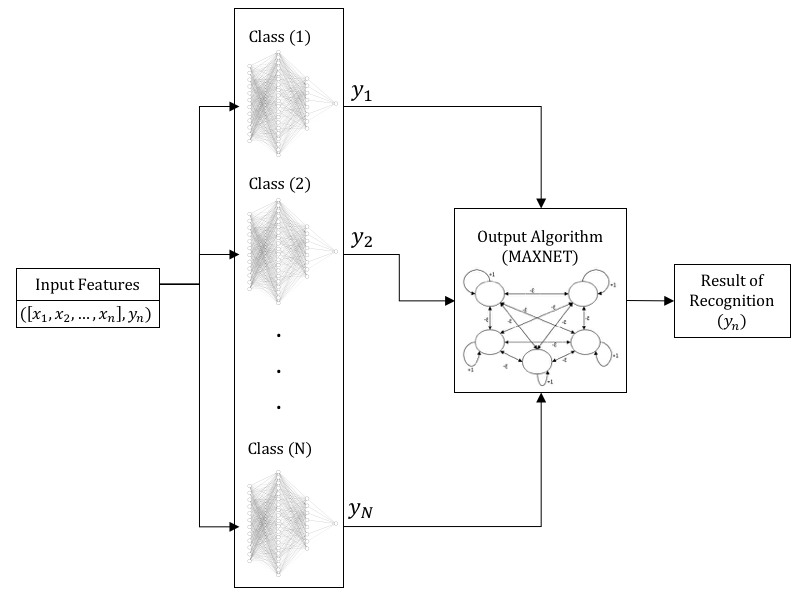}
\caption{The OCON model}
\label{ocon_model}
\end{figure}

Introduced in the '90s, the One-Class-One-Network (OCON) model \cite{10.1109/ISCAS.1991.176636} served as a solution for NNs \textit{parallel distributed processing}, aiming to overcome limitations of architectures that required full re-training when altering their dataset classes. Today the OCON resembles a simplified form of architecture \textit{ensembling} \ref{ocon_model}, where multiple complex networks are combined through other blocks or algorithms, to enhance the overall model accuracy. In the Anomaly Detection and Computer Vision fields \cite{MOYA1996463, DBLP:journals/corr/abs-1802-06360, 10.1109/TIP.2019.2917862}, the \textit{One-Class} approach consists of distributing a multi-output classification across a bank of independent sub-networks, each functioning as a \textit{context-specific} binary classifier. In our study, we divided a $12$-phoneme and $3$-genders classification tasks respectively into a bank of $12$ and $3$ independent and distributable classifiers, with identical architectural topology, aiming for an optimal \textit{average architecture estimation}.

If a discrete output label is needed, a context-specific output algorithm must be devised. While no literature references were found regarding OCON-specific output algorithms, figures in \cite{10.1109/ISCAS.1991.176636} suggest the involvement of a \textit{MaxNet} sub-network \cite{maxnet_paper}.

\begin{algorithm}[H]
\scriptsize
\caption{MaxNet Algorithm}
\label{maxnet}
\begin{algorithmic}
\Require $f(\cdot)$ \Comment{Activation function}
\Require $n$ \Comment{Nodes number}
\Require $\varepsilon \cong \frac{1}{n}$ \Comment{Inhibition magnitude}
\Require $\{y_1,\dots, y_n\}$ \Comment{Network outputs}
\Require \underline{criterion} \Comment{\textit{Winner-takes-all} evaluation}
\For {$k = (1,\dots, n)$} \do {} \Comment{Weights initialization loop}
    \If{$k = n$}
        \State $\theta_k = +1$ \Comment{Self-weight assignment}
    \Else
        \State $\theta_k=\varepsilon$ \Comment{Inhibition-weight assignment}
    \EndIf
\EndFor
\While {\underline{criterion}}
    \For {$k = (1, ..., n)$}
        \If{$i \neq j$}
            \State $y_k'=f(y_k-\theta_k\sum_{i=1}^{n}{y_i})$ \Comment{Competition}
        \Else
            \State $y_k'=y_k+\theta_k$ \Comment{Unitary increment}
        \EndIf
        \State $y_k \gets y_k'$ \Comment{New outputs assignment}
    \EndFor
\EndWhile
\end{algorithmic}
\end{algorithm}

The MaxNet can encounter critical flaws when multiple maxima occur in the input state, potentially leading to infinite \textit{competitive looping}. To mitigate this issue, the \textit{argument of the maxima} (\textit{ArgMax}) is employed, which typically returns a single value, representing the first occurrence of a maximum, when multiple exist. However, we find the classification \textit{logits} vector more beneficial for investigating phonetic and speakers class boundaries, features complexities and/or similarities.

\begin{figure}[!t]
\centering
\includegraphics[width=3.5in]{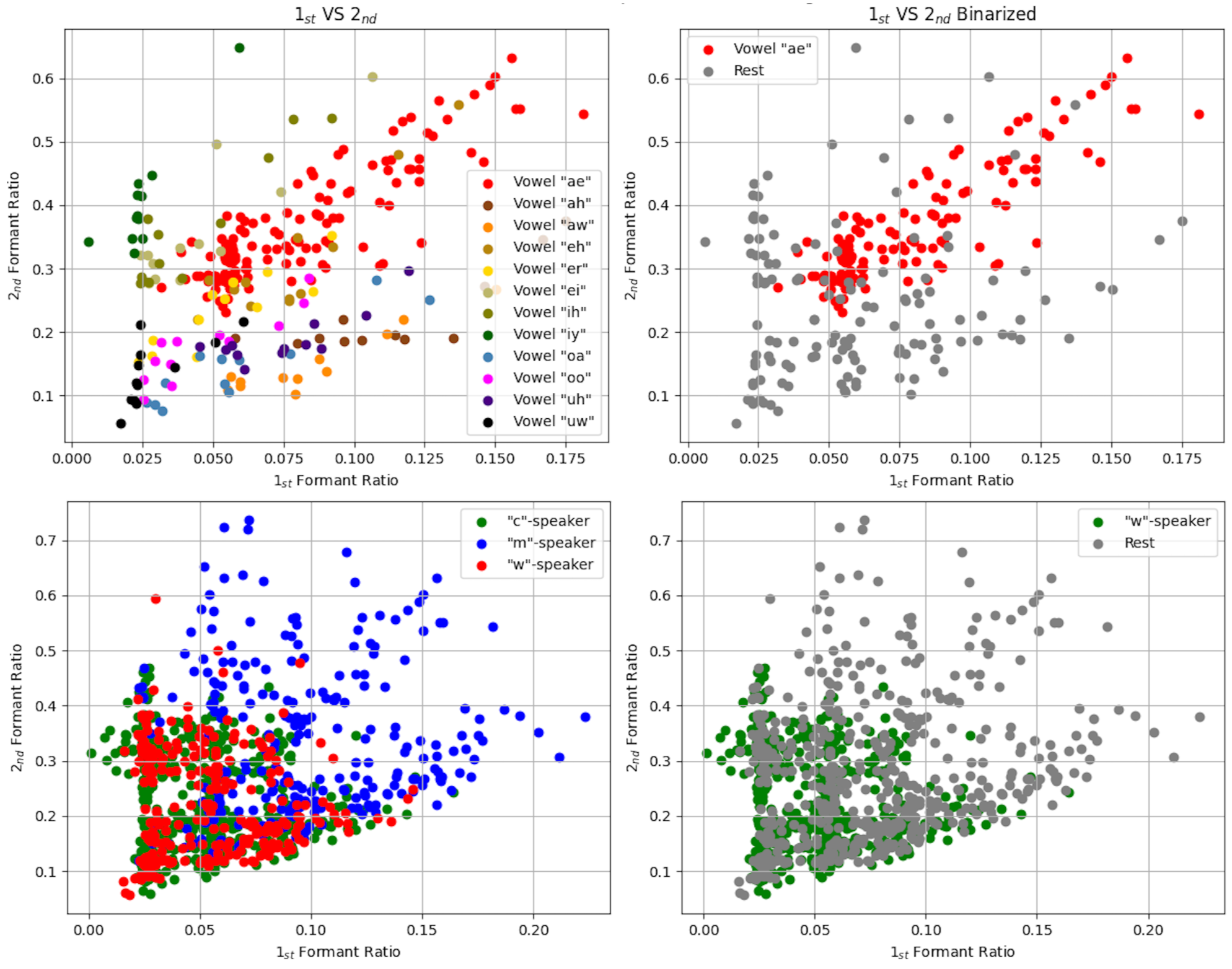}
\caption{HGCW dataset \textit{One-Hot encoding} examples}
\label{hgcw_one-hot}
\end{figure}

During supervised training, each sample label undergoes binarization (\textit{one-hot encoding}) specifically tailored to the One-class architecture, while features are concurrently fed into all classifiers. To automatize this process, we devised a custom one-hot encoding sub-routine (Alg.\ref{one-hot_enc}), so as to transform labels according to the incoming \texttt{True}-One-class. Additionally, to address classes under-representations (observed in Table \ref{hgcw_classes}), we perform a slight down-sampling of resulting training subsets.
\begin{algorithm}[H]
\scriptsize
\caption{HGCW One-Hot encoding}
\label{one-hot_enc}
\begin{algorithmic}
\Require $c$ \Comment{True-class index}
\Require $s$ \Comment{Phoneme groups size}
\Require $\mathcal{X}$ \Comment{Features dataset}
\Require $\mathcal{Y}$ \Comment{Dataset labels}
\State class$_1 =\mathcal{X}(c)$ \Comment{Initialize True-class subset}
\State size $=$ length$($class$_1)$ \Comment{Extract True-class size}
\State classes$_0 =$ list[ ] \Comment{Initialize False-class subsets}
\State sub-sizes $=$ $round(\frac{size}{11})$ \Comment{Compute False-classes size}
\For {$k$ in $\mathcal{Y}$} \do {} \Comment{Subsets selection loop}
    \If{$\mathcal{Y}_k \neq c$}
        \State class$_0 = rand(\mathcal{X}_k$, sub-sizes$)$ \Comment{Random downsampling}
        \State classes$_0$.append$($class$_0)$
    \Else
        \State pass
    \EndIf
\EndFor
\end{algorithmic}
\end{algorithm}
Alg.\ref{one-hot_enc} executes the one-hot encoding routine once per architecture training cycle, preceding the \textit{train-eval-test} splitting and the \textit{mini-batch} partitioning of features subsets. It ensures a balanced outliers quantity by allocating the same number of samples for all \texttt{False}-classes among the remaining $11$ (or $2$ for speakers), based on the available size of the \texttt{True}-class. If \texttt{True}-class sizes are not divisible by $11$, a variability of $1$ to $3$ samples is deemed acceptable. Speaker-based encoding for \textit{male}, \textit{female}, and \textit{children} classes is achievable (Fig. \ref{hgcw_one-hot}) in the same way (with less noticeable variability).

\subsection{Pseudo-NAS \& HPs-T search}
The term \textit{pseudo}-NAS, as discussed in Sec.\ref{subsec:architecture_and_model} and in \cite{Acce2406:OCON}, refers to the \textit{a prìori} constraint applied to the architecture topology (MLP). Our model evaluation will determine the optimal number of layers, and nodes per layer required to effectively address both phoneme and speakers gender recognition.

Conversely, \textit{grid-based} HPs search is a statistical method where all possible combinations of NNs HPs are independently sampled and evaluated through straightforward learning cycles. While theoretically effective, it can be a time-consuming solution due to the exponential increase in computational requirements (for narrowing resolutions): typically, all possible combinations must be tested before selecting the optimal one. We achieved a good trade-off by establishing independent resolutions for each HP beforehand, employing an \textit{informed} iterative approximation, summarized as follows:

(\emph{1}) define a specific subset of HPs (not necessarily all at once, potentially fixing others);

(\emph{2}) sample each HP with an arbitrary resolution;

(\emph{3}) test each combination of HPs and evaluate resulting \textit{temporary best estimates}. These can either serve as \textit{inheritable optimal estimates} for subsequent heuristic stages or guide parameter resolution sampling towards \textit{local good estimates}, in search of better sets;

(\emph{4}) repeat steps (\emph{2}) to (\emph{3}) as much as needed, to refine and improve the model configuration.

Acknowledging that this simplified approach roughly approximates theoretical grid-search, leading to potential misleading local minima in model costs, our goal remains to identify an average One-class topology in a computationally feasible manner.

Heuristic learning experiments involved partitioning the dataset into train ($70\%$), dev ($15\%$), and test ($15\%$) sets, with seeded initial states (for random initialization processes involved). Accuracy and mini-batch training times are measured, and results are averaged over a $3$-folded validation procedure for each One-class.
\begin{table}
\begin{center}
\caption{NAS \& Learning heuristic stage \\ IN = \textit{input nodes}, LR = \textit{learning rate}, HN/L = \textit{hidden nodes/layers}}
\label{experiment_1}
\begin{tabular}{| c | c | c |}
\hline
\textbf{Input Features} & \textbf{Fixed HPs}& \textbf{Testing HPs}\\
\hline
SS formant ratios & IN ($3$) & HN ($10, 50, 100$)\\
  & HL ($1$) & Backprop\\
  & activations (ReLU) & (\textit{Adam}, \textit{RMSProp})\\
  & states init. & LR\\
  & (\textit{standard} \cite{10.1109/ICCV.2015.123}, $b=0$) & ($10^{-3}, 10^{-4}, 10^{-5}$)\\
  & epochs ($1000$) & \\
  & batch size ($32$) & \\
  & k-folds ($3$)& \\
\hline
\textbf{TOT sets: 18} & \textbf{TOT architectures: 12} & \textbf{TOT cycles: 648}\\
\hline 
\end{tabular}
\end{center}
\end{table}

\begin{figure}[!t]
\centering
\includegraphics[width=3.4in]{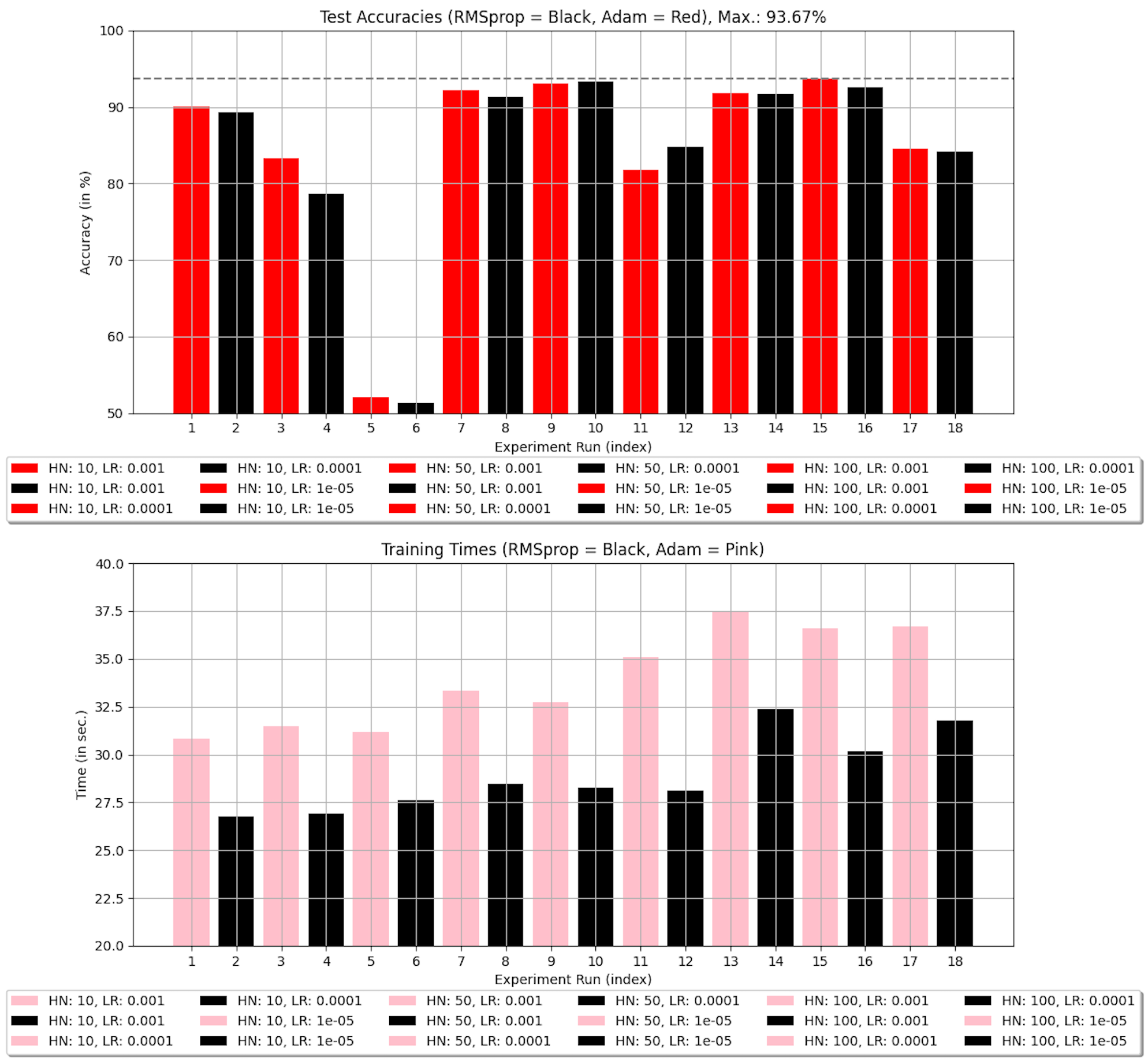}
\caption{$1st$ heuristic stage results}
\label{experiment_1_acc_times}
\end{figure}

In the first \textit{architectural} heuristic stage (Table \ref{experiment_1}), two combinations ($10$th and $15$th) yielded similar average accuracies ($93.67\%$, Fig.\ref{experiment_1_acc_times}). The RMSProp optimizer \cite{Tieleman} demonstrated better mitigation of the increasing trend in learning times, compared to Adam \cite{Kingma}. However, we opted for the top-performing setup: HL: $1$, HN=$100$, LR=$10^{-4}$, Backprop: Adam.

Following heuristic stages were designed to assess the incremental introduction of \textit{regularization} techniques, and to evaluate potential advantages.
\begin{table}
\begin{center}
\caption{$2^{nd}, 3^{rd}$ \& $4^{th}$ heuristic stage (HP-T Regularization)\\ IN = \textit{input nodes}, LR = \textit{learning rate}, HN/L = \textit{hidden nodes/layers}}
\label{experiment_reg}
\begin{tabular}{| c | c | c |}
\hline
\textbf{DropOut HPs}& \textbf{Batch-norm HPs}& \textbf{L2-Norm HPs}\\
\hline
IN DropOut rate & LR & L2-Norm\\
(0.8, 0.9) & ($10^{-3}$, $10^{-4}$, $10^{-5}$) & $\lambda (10^{-2}, 10^{-3}, 10^{-4})$\\
HN DropOut rate & Batch-Norm & \\
($[0.5, 1.]$, res.: $0.1$) & & \\
\hline 
LR ($10^{-4}$) &  & LR ($10^{-4}$)\\
k-folds ($6$) & k-folds ($10$) & k-folds ($10$) \\
& batch size (32) & batch size (32) \\
epochs (3000) & epochs (1000) & epochs (1000) \\
\hline
\textbf{TOT cycles: 864} & \textbf{TOT cycles: 360} & \textbf{TOT cycles: 360}\\
\hline 
\end{tabular}
\end{center}
\end{table}
\begin{figure}[!t]
\centering
\includegraphics[width=3.5in]{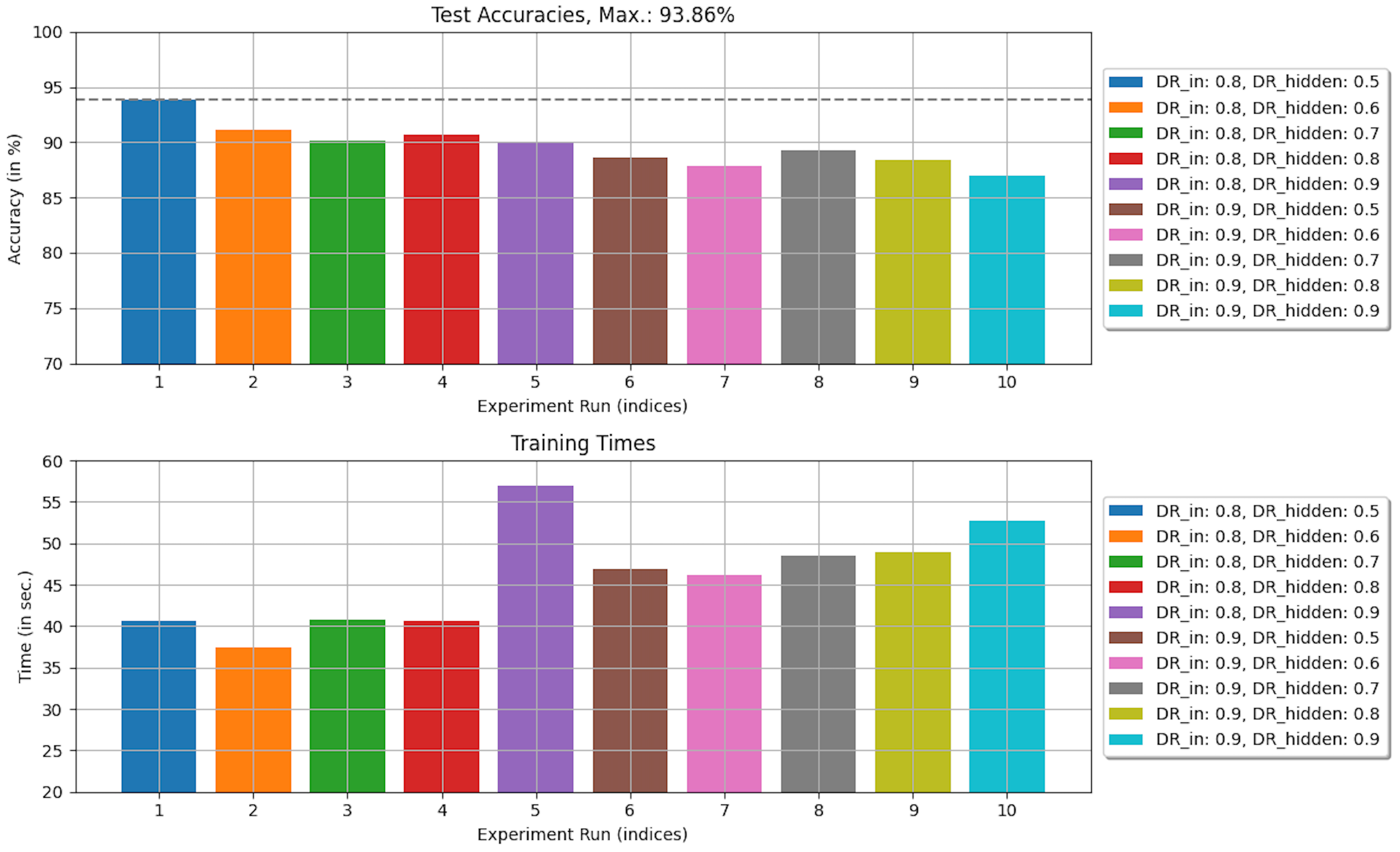}
\caption{$2^{nd}$ heuristic stage results (Dropout)}
\label{experiment_2_acc_times}
\end{figure}

In DropOut tests \cite{10.5555/2627435.2670313} (Table \ref{experiment_reg}, Fig.\ref{experiment_2_acc_times}), we observed that the fastest run ($1$st) also reached the highest accuracy. We achieved a $+0.19\%$ accuracy, at the expense of $+3.4$sec. in the average training time, with DropOut probabilities set to $80\%$ for input nodes and $50\%$ for hidden nodes.
\begin{figure}[!t]
\centering
\includegraphics[width=3.5in]{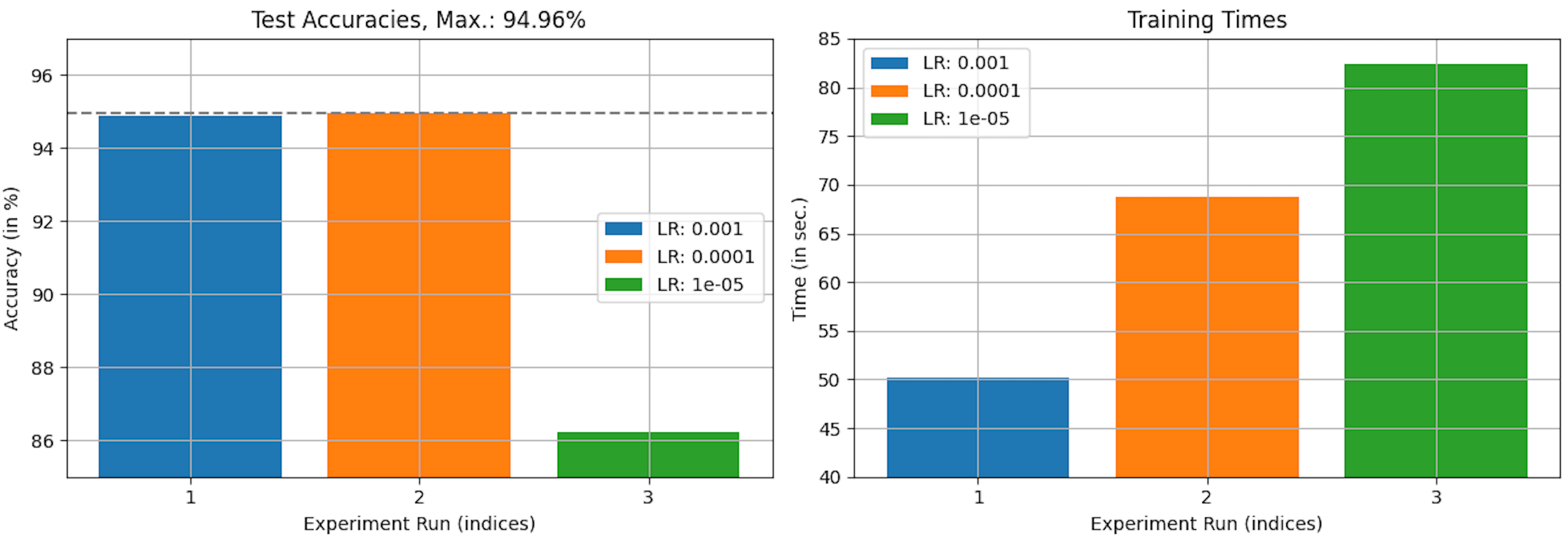}
\caption{$3^{rd}$ heuristic stage results (Batch-Norm)}
\label{experiment_3_acc_times}
\end{figure}

In Batch-norm tests \cite{10.5555/3045118.3045167}, after re-evaluating LRs, it was confirmed that LR$=10^{-4}$ yielded the best results: a significant $+1.1\%$ in test accuracy, despite nearly doubling average training times.

\begin{figure}[!t]
\centering
\includegraphics[width=3.5in]{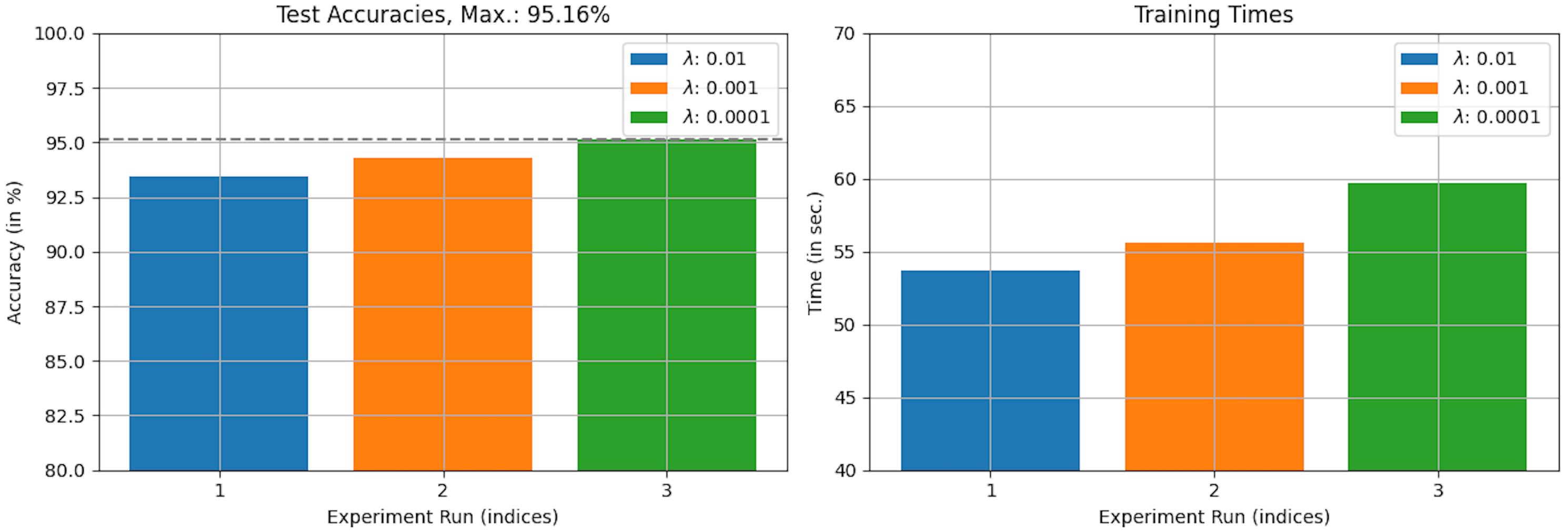}
\caption{$4^{th}$ heuristic stage results (L2-Norm)}
\label{experiment_4_acc_times}
\end{figure}

For L$2$-Norm tests (Table \ref{experiment_reg}) (also Ridge penalty \cite{10.2307/1271436}), we found an optimal $\lambda$ (weight decay) of $10^{-4}$ (Fig.\ref{experiment_4_acc_times}), resulting in a $+0.19\%$ for the average accuracy and a decrease in the average training time, now below $60$ seconds.
Table \ref{one_class} illustrates our overall averaged One-Class proposal.
\begin{table}
\begin{center}
\caption{One-Class architecture (MLP)\\ IL = \textit{input layer}, LR = \textit{learning rate}, HN/L = \textit{hidden nodes/layers},\\ ON = \textit{output nodes}}
\label{one_class}
\begin{tabular}{| c | c | c |}
\hline
\textbf{Architecture} & \textbf{Features} & \textbf{Learning}\\
\hline
IL: $3$ nodes & $\omega$ init.: \textit{Kaiming-He} norm. & \textit{Adam} optim.\\
HL: 1, $100$ nodes  & $b$ init.: $0$ & LR: $10^{-4}$\\
ON: 1 (\textit{logit}) & One-hot encoding & Mini-bacth\\
ReLU (common)& (Dataset \textit{re-shuffling}) & ($32$ samples)\\
\hline
IN-DR: $0.8$ & Batch-Norm  & L2-$\lambda$: $10^{-4}$\\
HL-DR: $0.5$ & & \\
\hline
\end{tabular}
\end{center}
\end{table}

\section{Model Training \& Results Discussion}
A parallelized set of independently trainable One-Class architectures was implemented and trained using CPU runtimes (on Google Colab) to efficiently measure isolated training cycle performances and resource consumptions. We stress that the OCON architecture learning relies on the backpropagation loop of each MLP. During inference, it involves extracting sample features, computing $12$ parallel one-hot encodings, and performing an ArgMax search to determine the maximum value (predicted label) within the $12$-logit probabilities vector. 
Following this, we conducted phoneme recognition experiments to evaluate the efficiency of each dataset sub-structure (Sec.\ref{features_sec}). An Early-Stopping training strategy \cite{bai} was adopted, incorporating a two-variable escape condition: a minimum loss threshold (averaging among the last $50$ training samples' loss) and a minimum test accuracy threshold based on the last batch results. These variables were further empirically assessed to ensure practical convergence of training cycles, with each cycle not exceeding a maximum amount of $25$-$30$ minutes. While the learning phases may not be fully optimized, they were deemed satisfactory for the purposes of our study.

\subsection{Phonemes recognition}
\begin{figure}[!t]
\centering
\includegraphics[width=3.5in]{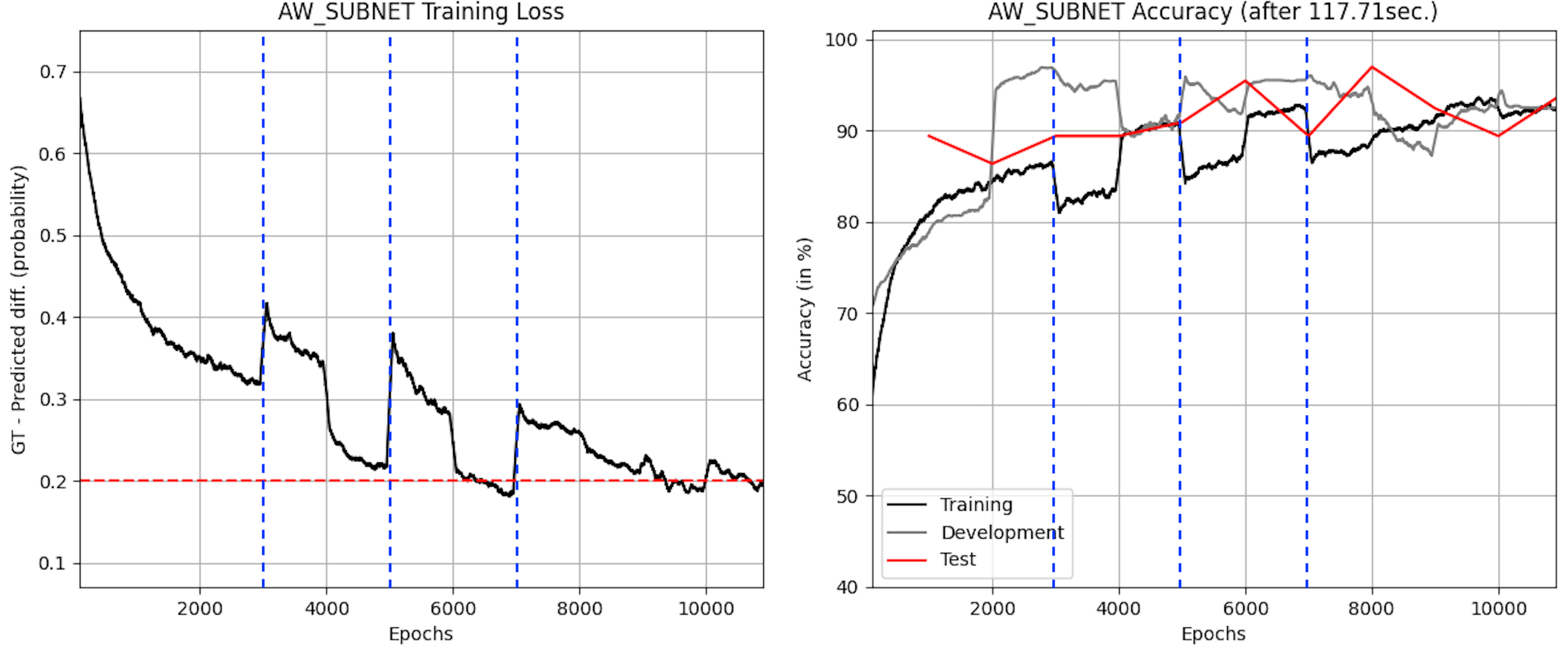}
\caption{Early-stopping spike examples}
\label{spikes}
\end{figure}

In \cite{Acce2406:OCON} we evaluated the OCON model using the steady-state (SS) dataset variant (Table \ref{phonemes_1}):
\begin{table}
\begin{center}
\caption{$1^{st}$ Experiment: \textit{SS}-phonetic classification}
\label{phonemes_1}
\begin{tabular}{| c | c | c |}
\hline
\textbf{Features} & \textbf{Training} & \textbf{Early-Stopping}\\
\hline
SS formant ratios & epochs: $1000$& Loss thresh.: $0.2$\\
 & (for each \textit{batch-set}) & Accuracy thresh.: $90$\%\\
 & Re-shuffling & \\
 & balancing tol.: $0.01$ & \\
\hline
\textbf{Phonemes} & \textbf{Test Accuracy} (\%) & \textbf{Training times} (sec.)\\
\hline
ae & $86.27$ & $247.62$\\
ah & $90.85$ & $85.67$\\
aw & $86.09$ & $117.71$\\
eh & $89.05$ & $345.20$\\
er & $91.90$ & $25.71$\\
ei & $84.97$ & $539.79$\\
ih & $87.38$ & $207.92$\\
iy & $92.21$ & $33.78$\\
oa & $82.31$ & $120.20$\\
oo & $85.96$ & $396.59$\\
uh & $85.65$ & $485.34$\\
uw & $90.91$ & $219.23$\\
\hline
\textbf{OCON Acc.:} $70\%$ & \textbf{AVG Acc.:} $87.79\%$ & \textbf{AVG Time:} $235.40$sec.\\
\hline
\end{tabular}
\end{center}
\end{table}
training revealed that several loss functions and training accuracy curves visibly plateaued, punctuated by periodic \textit{spikes} indicating instances of consistent-learning batch re-shuffling (Fig.\ref{spikes}). Interestingly, the \textit{er} and \textit{iy} phoneme classes exhibited substantial representation, showing almost no changes (in curve trends) post-encoding or re-shuffling (Table \ref{phonemes_1}).

Overall accuracies were computed using a binary threshold of $0.5$ across the entire dataset classes. While certain MLPs effectively segregated probabilities, notable errors persisted between phonemically (\textit{aural}) similar classes, such as \textit{ae} and \textit{eh}, and \textit{er} and \textit{ei}.

Hidden dataset biases, such as similarities in formantic disposition between \textit{children} and \textit{women} utterances, were re-examined and filtering out these biases led to slight improvements in class boundaries separation, despite increasing training loops duration. Attempts to enhance speakers gender boundaries by re-introducing $F0$s data, proved to be unsuccessful (AVG acc.: $88.80\%$, OCON acc.: $74\%$).

The most effective feature set consisted of $4$ temporal tracks of $3$ formant ratios (Table \ref{phonemes_3}), which significantly boosted accuracy, reduced training times, and mitigated side effects of Early-Stopping, approaching the accuracy goal referenced in \cite{0.1088/1742-6596/2466/1/012008} of $90\%$ (Table \ref{roc_metrics}).
\begin{table}
\begin{center}
\caption{$3^{rd}$ Experiment: \textit{Time-Tracks}-phonetic classification}
\label{phonemes_3}
\begin{tabular}{| c | c | c |}
\hline
\textbf{Features} & \textbf{Training} & \textbf{Early-Stopping}\\
\hline
$10$\%, $50$\% & epochs: $1000$& Loss thresh.: $0.15$\\
SS, $80$\% & (for each \textit{batch-set}) & Accuracy thresh.: $95$\%\\
formant ratios & Re-shuffling & \\
 & balancing tol.: $0.01$ & \\
\hline
\textbf{Phonemes} & \textbf{Test Accuracy} (\%) & \textbf{Training times} (sec.)\\
\hline
ae & $94.55$ & $72.17$\\
ah & $91.80$ & $156.37$\\
aw & $89.86$ & $71.47$\\
eh & $93.74$ & $540.39$\\
er & $93.43$ & $28.05$\\
ei & $96.37$ & $104.98$\\
ih & $94.55$ & $97.20$\\
iy & $96.49$ & $38.59$\\
oa & $93.49$ & $49.43$\\
oo & $95.62$ & $96.17$\\
uh & $90.98$ & $649$\\
uw & $93.74$ & $108.30$\\
\hline
\textbf{OCON Acc.:} $90\%$ & \textbf{AVG Acc.:} $93.72\%$ & \textbf{AVG Time:} $167.68$sec.\\
\hline
\end{tabular}
\end{center}
\end{table}

\subsection{Speaker recognition}
We aim to determine the minimum amount of formant features required for identifying speakers' gender with our best model. The overall architecture was simplified to $3$x One-Classes (Table \ref{speakers}): \textit{men}, \textit{women} and \textit{children}, with normalized $F0$s reintroduced in the input set, to improve classes separability.
\begin{table}
\begin{center}
\caption{\textit{Time-Tracks}-speaker classification}
\label{speakers}
\begin{tabular}{| c | c | c |}
\hline
\textbf{Features} & \textbf{Training} & \textbf{Early-Stopping}\\
\hline
$10$\%, $50$\% & epochs: 1000 & Loss thresh.:\\
SS, $80$\% & (for each \textit{batch-set}) & $0.36, 0.08, 0.45$\\
formant ratios & Re-shuffling & Accuracy thresh.:\\
+ \textit{min-max}ed $F0$s & balancing tol.: $0.01$ & $80\%, 97\%, 80\%$\%\\
\hline
\textbf{Speakers} & \textbf{Test Accuracy} (\%) & \textbf{Training times} (sec.)\\
\hline
children & $82.03$ & $310.26$\\
men & $97.75$ & $154.15$\\
women & $75.70$ & $503.28$\\
\hline
\textbf{OCON Acc.:} $80\%$ & \textbf{AVG Acc.:} $85.15\%$ & \textbf{AVG Time:} $322.56$sec.\\
\hline
\end{tabular}
\end{center}
\end{table}
Class-dependent Early-stopping criteria were defined due to the significant amount of adjustments required for proper training convergence: $0.36$, $0.08$, $0.45$ loss thresholds, $80\%$, $97\%$, $80\%$ accuracy thresholds (respectively for \textit{children, male} and \textit{women}).

\textit{Women} and \textit{children} classes faced challenges in loss minimization while the \textit{men} MLP converged rapidly to low error rates (almost $100$\% of accuracy). These results suggest better class representation for \textit{men} and confirmed known difficulties in aural partitioning between \textit{children} and certain adult \textit{female} voices (\textit{aural} similarities). Evaluation conducted upon the entire dataset's inference revealed lower False Positives (FPs) for the \textit{male} class and higher FP rates for \textit{children} and \textit{women} inferences.

The OCON model achieved a speaker genders recognition accuracy between $80\%$ to $85\%$, suggesting potential increasing reliability according to higher time-tracks number (more than $3$x formant ratios, per speaker). To finalize the statistical overview \cite{10.1016/j.aci.2018.08.003, 10.48550/arXiv.2010.16061} we provide related \textit{confusion matrices} (Tables \ref{acc_metrics_1}), \textit{Receiver Operating Characteristics} (ROC) curves analysis, \textit{Area-Under-the-Curve} (AUC) computations and \textit{Detection Error Tradeoffs} (DET) evaluations \cite{10.48550/arXiv.2010.16061, 10.1007/s10664-020-09861-4, 10.1109/ICASSP.2007.367205}: see related notebooks in our \href{https://github.com/StefanoGiacomelli/Vowel_phonemes_Analysis_and_Classification_by_means_of_OCON_rectifiers_Deep_Learning_Architectures}{GitHub repository}.
\begin{table}
\begin{center}
\caption{OCON Normalized Accuracy Metrics}
\label{acc_metrics_1}
\begin{tabular}{| c | c | c | c | c |}
\hline
\textbf{One-Class} & \textbf{Accuracy} & \textbf{Precision} & \textbf{Recall} & \textbf{F1-Score}\\
\hline
ae & $0.9737$ & $0.9568$ & $0.9925$ & $0.9744$\\
ah & $0.9625$ & $0.9310$ & $1.0000$ & $0.9643$\\
aw & $0.9509$ & $0.9110$ & $1.0000$ & $0.9534$\\
eh & $0.9631$ & $0.9388$ & $0.9928$ & $0.9650$\\
er & $0.9605$ & $0.9291$ & $1.0000$ & $0.9633$\\
ei & $0.9717$ & $0.9542$ & $0.9921$ & $0.9728$\\
ih & $0.9742$ & $0.9521$ & $1.0000$ & $0.9754$\\
iy & $0.9837$ & $0.9688$ & $1.0000$ & $0.9841$\\
oa & $0.9664$ & $0.9379$ & $1.0000$ & $0.9680$\\
oo & $0.9852$ & $0.9720$ & $1.0000$ & $0.9858$\\
uh & $0.9593$ & $0.9262$ & $1.0000$ & $0.9617$\\
uw & $0.9664$ & $0.9441$ & $0.9926$ & $0.9677$\\
\hline
\hline
\textbf{One-Class} & \textbf{TPs} & \textbf{FPs} & \textbf{FNs} & \textbf{TNs}\\
\hline
ae & $133$ & $6$ & $1$ & $126$\\
ah & $135$ & $10$ & $0$ & $122$\\
aw & $133$ & $13$ & $0$ & $119$\\
eh & $138$ & $9$ & $1$ & $123$\\
er & $118$ & $9$ & $0$ & $101$\\
ei & $125$ & $6$ & $1$ & $115$\\
ih & $139$ & $7$ & $0$ & $125$\\
iy & $124$ & $4$ & $0$ & $117$\\
oa & $136$ & $9$ & $0$ & $123$\\
oo & $139$ & $4$ & $0$ & $128$\\
uh & $138$ & $11$ & $0$ & $121$\\
uw & $135$ & $8$ & $1$ & $124$\\
\hline
\end{tabular}
\end{center}
\end{table}

\begin{table}
\begin{center}
\caption{OCON Normalized ROC-AUC/DET metrics}
\label{roc_metrics}
\begin{tabular}{| c | c | c | c | c | c |}
\hline
\textbf{One-Class} & \textbf{ER} & \textbf{FDR} & \textbf{FOR} & \textbf{NPV} & \textbf{AUC}\\
\hline
ae & $0.02$ & $0.03$ & $0.01$ & $0.99$ & $0.9986$\\
ah & $0.03$ & $0.06$ & $0.00$ & $1.00$ & $0.9866$\\
aw & $0.03$ & $0.06$ & $0.00$ & $1.00$ & $0.9980$\\
eh & $0.02$ & $0.03$ & $0.01$ & $0.99$ & $0.9934$\\
er & $0.02$ & $0.03$ & $0.00$ & $1.00$ & $0.9935$\\
ei & $0.03$ & $0.05$ & $0.01$ & $0.99$ & $0.9979$\\
ih & $0.03$ & $0.05$ & $0.00$ & $1.00$ & $0.9996$\\
iy & $0.01$ & $0.02$ & $0.00$ & $1.00$ & $0.9994$\\
oa & $0.04$ & $0.07$ & $0.00$ & $1.00$ & $0.9898$\\
oo & $0.01$ & $0.01$ & $0.00$ & $1.00$ & $1.0000$\\
uh & $0.03$ & $0.05$ & $0.00$ & $1.00$ & $0.9950$\\
uw & $0.03$ & $0.06$ & $0.01$ & $0.99$ & $0.9965$\\
\hline
\end{tabular}
\end{center}
\end{table}

\subsection{Energy efficiency}
\begin{figure}[!t]
\centering
\includegraphics[width=2in]{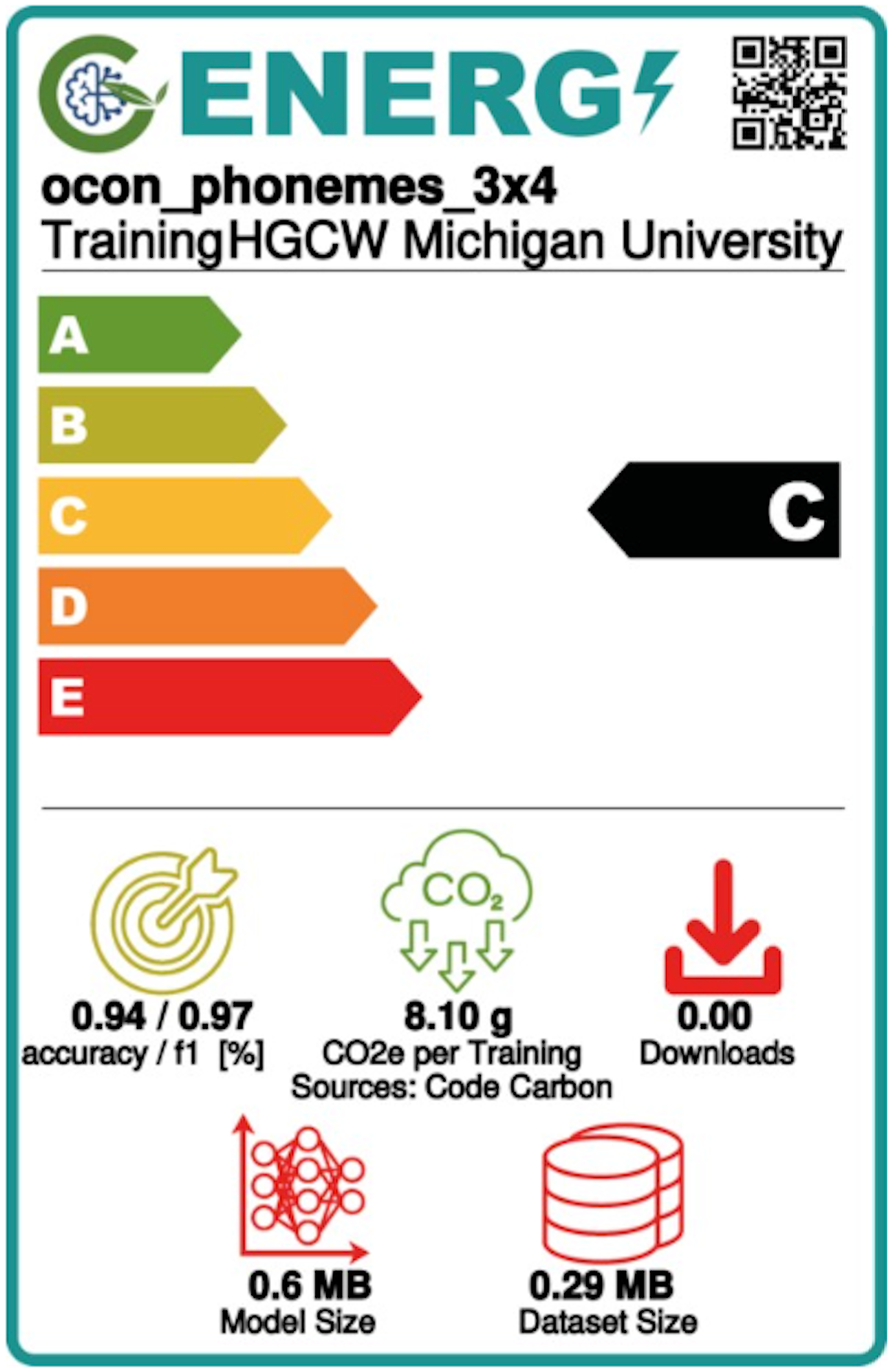}
\caption{OCON Energy label}
\label{energy_label}
\end{figure}

We focused on experimental sustainability grabbing insights from the Green-AI field \cite{Henderson, 10.1145/3381831}. Using \texttt{CodeCarbon}, a custom Python-API for \texttt{Intel-RAPL} and \texttt{Nvidia-smi} libraries, we tracked CPU, disk, and RAM usage during model training. Energy metrics and estimated CO2 emissions were recorded in a \texttt{.csv} file and analyzed through a \href{http://energy-label.streamlit.app}{web-based applet} developed by the GESSI research group (Universitat Politècnica de Catalunya), as part of the GAISSA research project. This analysis provided efficiency and accuracy labels (Fig.\ref{energy_label}) via HuggingFace database comparison, indicating strong sustainability of our pseudo-NAS approach, without compromising resulting accuracies.

The entire speakers task training cycle required approximately $36$ minutes, with an average consumption of $42.5$W for CPU and $4.27$W for RAM. Carbon dioxide emissions (CO$_2$eq) were estimated as the product between grams of CO$_2$ emitted per KW-hour of electricity ($0.025$KW/h for CPU, $0.003$KW/h for RAM) and the energy consumed by the computational infrastructure: resulting in $0.008$Kg, with an emission rate of $3.75 \times 10^{-6}$Kg/s.
\begin{table}
\begin{center}
\caption{OCON model Energy profile}
\label{energy_metrics}
\begin{tabular}{| c | c |}
\hline
\textbf{Feature} & \textbf{Value}\\
\hline
TOT Parameters & $140412$ ($11701$ each)\\
Estimated Size & $0.6$MB ($0.05$ each)\\
TOT mul-adds & $1.2 \cdot 10^{5}$\\
Dataset size & $287$KB (compressed)\\
Energy Measurement Date & 2023-12-04\\
Energy Measurement Time & 10:51:38\\
\hline
Profiling Software & \texttt{CodeCarbon}\\
Emissions & $0.0081$KgCO2\\
Emission rate & $3.7453 \cdot 10^{-6}$KgCO2h\\
\hline
Training Architecture & CPU (x2)\\
Model & Intel(R) Xeon(R) $2.20$GHz\\
Cache size & $56320$KB\\
Power & $42.5$W\\
Energy & $0.0255$kWh\\
TOT RAM & $12.67$GB\\
RAM Power & $4.7543$W\\
RAM Energy & $0.0028$kWh\\
TOT Energy consumed & $0.0284$kWh\\
\hline
Cloud Service & Google Colab\\
Server Location & South Carolina (USA)\\
OS & \texttt{Linux5.15.120x86\_64 glibc2.35}\\
Python & \texttt{3.10.12}\\
Ext. packages & \texttt{NumPy, MatPlotLib, PyTorch,}\\
 & \texttt{SciPy, SKlearn}\\
\hline
\end{tabular}
\end{center}
\end{table}

\section{Conclusions and future works}
We are aware that a single Perceptron can easily predict speech signal samples approximating LPA results \cite{10.1109/ICNN.1995.487832}. Our model proposal can therefore be seen as an ad-hoc integration \textit{head} for a complex Perceptron-based \textit{formant neural framework}. Despite the active research on formant estimation leveraging convolutional and recurrent layers (backbone stages) \cite{ALKU2023101515, 10.1121/1.5088048}, we believe that our approach, employing pseudo-NAS/HP-T techniques entirely scripted and executed on Colab free-tier notebooks, can be broadly re-applied to effectively evaluate the efficiency of newer SoA CNNs building blocks (lihtweight-CNNs) in terms of parameters reduction and computational complexity.

Our foundational research model demonstrates high distributability, with each classifier independently re-trainable and sufficiently lightweight for constrained computational contexts (or hardwares), making it suitable for integration into pre-existent complex architectures and on-board sensory constrained hardware. Optimization techniques such as parameters \textit{pruning} and \textit{quantization} could further improve its memory consumption at inference time. Additionally, its modular structure allows for easy adaptation to different language and speakers grouping contexts (being more LGBTQIA+ friendly, despite the use of simple binary-labeled dataset). A pre-compiled \texttt{TorchScript} version of our classification stages, successfully run on testbed E2E sensing and processing devices (kindly provided by the developer company). 

We try to challenge the notion that larger (and heterogeneous) datasets or complex (Transformer-based) models inherently yield better accuracies: asserting instead that our approach offers good generalizability and adaptability, despite known limitations in training sample size. While we encountered difficulty in finding extensive pre-processed datasets, we will re-validate our findings by expanding our dataset sources, potentially validating TI-MIT, UCLAPhoneticsSet and AudioSet.

Our proposal for linear features processing confirms that altering speech signal spectra in non-linear auditory-based ways it's not always optimal for \textit{descriptive} speech modeling. However, we intend to reconsider solutions proposed in the existing literature.

Regarding sustainability, we're pleased to find that the CO$2$e emissions for fully retraining our model are just over half the emissions from the entire lifecycle of a \textit{single cigarette}. 

Future research could explore enhancing label (class) selection by applying training assurance scaling coefficients to output One-Class probabilities, aiming to increase model \textit{reliability}. This approach involves analyzing epochs during which the classifier maintains the loss below specified Early-stopping thresholds. Further refinement of output probabilities could utilize derivatives of the loss curve, particularly useful in cases of too rapid training error minimization.

\section*{Acknowledgments}
\noindent We extend our sincere thanks to the Electronic Music Department at A. Casella Conservatory (L'Aquila) for their invaluable assistance in developing this project. ChatGPT (OpenAI) was used here to support typos correction and text formatting.

\bibliographystyle{IEEEtran}
\bibliography{References.bib}

\end{document}